\documentclass[aps,pra,amsfonts,reprint,superscriptaddress,footinbib]{revtex4-1}

\usepackage{amsmath}
\usepackage{amstext, amssymb}
\usepackage[normalem]{ulem}
\usepackage{bbold}
\usepackage{graphicx}
\usepackage{color}

\newcommand{\comment}[1]{}

\newcommand{\id}{\mathbb{1}}

\begin{document}

\title{Asymptotic Floquet states of non-Markovian systems}

\author{Luca Magazz\`u}

\affiliation{Institute of Physics, University of Augsburg, Universit\"atsstrasse 1, D-86135 Augsburg, Germany}
\author{Sergey Denisov}
\affiliation{Institute of Physics, University of Augsburg, Universit\"atsstrasse 1, D-86135 Augsburg, Germany}
\affiliation{Department of Applied Mathematics, Lobachevsky State University of Nizhny Novgorod, Nizhny Novgorod 603950, Russia}
\affiliation{Nanosystems Initiative Munich, Schellingstra{\ss}e 4, D-80799 M\"unchen, Germany}
\author{Peter H\"anggi}
\affiliation{Institute of Physics, University of Augsburg, Universit\"atsstrasse 1, D-86135 Augsburg, Germany}
\affiliation{Department of Applied Mathematics, Lobachevsky State University of Nizhny Novgorod, Nizhny Novgorod 603950, Russia}
\affiliation{Nanosystems Initiative Munich, Schellingstra{\ss}e 4, D-80799 M\"unchen, Germany}

\date{\today}

\begin{abstract}
We propose a method to find asymptotic states of a class of periodically modulated open systems 
which are outside the range of validity of the Floquet theory due to the presence of memory effects. 
The method is based on a Floquet treatment of the time-local, memoryless dynamics taking place in a minimally enlarged state space where the original system is coupled 
to auxiliary -- typically non-physical -- variables.
A projection of the Floquet  solution  into the physical subspace returns the sought asymptotic state of the  system. 
The spectral gap of the Floquet propagator acting in the enlarged state space can be  used to estimate the relaxation time.
We illustrate the method with a  modulated version of quantum random walk model.
\end{abstract}


\maketitle

\section{Introduction}
Periodically driven systems  can 
exhibit a spectrum of states which are 
unattainable in the static limit. 
This makes the idea of modulations appealing to various fields,
ranging from dynamical chaos theory \cite{Kadanoff1993} and chemical kinetics~\cite{Petrov1997} to neuroscience~\cite{Herrmann2001} 
and quantum physics ~\cite{Shirley1965,Sambe1973,Grifoni1998,Bukov2015,Eckardt2015}. 
In the latter field, periodic driving was  used to realize new topological states~\cite{Lindner2011,Liu2013}, 
engineer artificial gauge fields~\cite{Struck2011,Goldman2014}, and create 
so-called `Floquet time crystals'  \cite{Else2016,Zhang2017,Choi2017}. \\
\indent Typically, one models  periodically modulated systems via  linear differential
equations  with time-periodic coefficients whose solution is provided by Floquet theory~\cite{Floquet1883,Yakubovich1975}.
The key prerequisite to construct such a model is the time-local character of the system dynamics which means 
that the future of the system depends on its current state  and not on its history. For a
system with time-local and contractive (in terms of some proper norm) dynamics, the fate of the system  is specified by the asymptotic state(s).
A periodically modulated system  interacting with a broadband environment in the Markovian limit, evolves towards an asymptotic state which 
is periodic with the period of the modulations~\cite{Jung1993,Gammaitoni1998,Hartmann2017}.
On the model level, this state represents a limit cycle solution of the dissipative equations describing the system dynamics \cite{Yakubovich1975}.
Very recently, an idea to combine modulations and dissipation  to explore  many-body quantum states \cite{Vorberg2013,Lazarides2017} has emerged
as a natural  extension of the established Hamiltonian-oriented approach~\cite{Goldman2014,Bukov2015,Sommer2016,Meinert2016,Eckardt2017}.
Still, the use of Floquet theory in the dissipative context implies that the  equations used to describe the dynamics remain local in time~\cite{Kohler2005,Alicki2007,Bastidas2017}.\\
\indent What are  the  asymptotic Floquet  states of systems governed by time-\textit{non}local evolution equations?
This question is of a special relevance in the context of quantum non-Markovian dynamics, a topic being actively explored now both in theoretical \cite{Breuer2016} and experimental \cite{Liu2011} domains. 
Here we present a method to get the answer for a broad class of periodically modulated systems whose evolution is governed by memory-kernel master equations.
We show that the corresponding asymptotic solutions possess the periodicity of the modulations and have the form prescribed by the recently introduced generalized Floquet theorem (gFT)~\cite{Traversa2013}.
\section{Method}
 We consider physical systems, both classical and quantum, whose state is described by a $n$-dimensional vector $\mathbf{x}(t)$ obeying a generalized master equation of the form
\begin{equation}\label{MKME}
\dot{\mathbf{x}}(t) =\int_{t_0}^{t}dt'\mathbf{K}(t,t')\mathbf{x}(t')+\mathbf{z}(t)\;,
\end{equation}
with an integrable memory kernel (MK) matrix $\mathbf{K}(t,t')$ and an asymptotically vanishing inhomogeneous 
term, $\lim_{t \rightarrow \infty} \mathbf{z}(t) =  \mathbf{0}$~\footnote{For $\lim_{t \rightarrow \infty}\mathbf{z}(t+T) = \mathbf{z}(t)\neq \mathbf{0}$,
the method has to be complemented with a non-homogeneous extension of Floquet theory \cite{Yakubovich1975}.}. 
The vector $\mathbf{x}(t)$ may describe, for example, a  $n$-state classical system or the density 
matrix of a quantum system in a suitable representation~\cite{Vacchini2013,Vacchini2016}. 
We assume the 
MK to be biperiodic, i.e., $\mathbf{K}(t+T,t'+T) =\mathbf{K}(t,t')$,  where $T$ denotes the period of the modulation. The above  
requirements ensure that, in the limit $t\rightarrow \infty$,  the action 
of the operator $\mathcal{L}_t       \{ \mathbf{x}(t) \}=\int_{t_0}^{t}dt'\mathbf{K}(t,t')\mathbf{x}(t')+\mathbf{z}(t)$ commutes 
with that of the one-period translation operator  $\mathcal{S}_{T}\{\mathbf{x}(t)\}=\mathbf{x}(t+T)$, thus entailing the applicability 
of the gFT to the asymptotic dynamics induced by Eq.~(\ref{MKME})~\cite{note1}. 
Note that the condition of biperiodicity is automatically satisfied if the MK depends 
exclusively on the difference $\tau=t-t'$ or if the MK is periodic with respect to $t$ ($t'$) and depends only on $t$ ($t'$) and $\tau$. 
\\
\indent Our approach to determine the asymptotic solution of Eq.~(\ref{MKME}) 
is based on the idea of embedding the system  into an enlarged state 
space where the physical variable $\mathbf{x}(t)$ is coupled to 
an auxiliary vector variable $\mathbf{u}(t)$. The coupling is realized in such a way that the resulting 
extended system described by the new vector $\mathbf{v}^{\rm T}(t) = [\mathbf{x}^{\rm T}(t), \mathbf{u}^{\rm T}(t)]$ (concatenation) 
obeys a time-local equation satisfying the conditions of the standard Floquet theorem. The latter provides the solution for 
the state of the extended vector $\mathbf{v}(t)$ whose projection into the physical $\mathbf{x}$-subspace constitutes the asymptotic 
Floquet state of the system.

The concept of embedding is well known within the theory of classical stochastic processes
\cite{Grabert1977,Grabert1978,
Grabert1980,Grigolini1982,
Kupferman2004,Siegle2010} where it is related to celebrated  Erlang's method of stages \cite{Cox1977}.
Embedding schemes are also employed~\cite{Breuer2004,Budini2013,Kretschmer2016} in the recently established field of quantum non-Markovian processes~\cite{Breuer2008,Vacchini2016,Breuer2016}. 
However, in the above cases the embedding is \textit{physical}, i.e.,  
the new auxiliary variables (states or degrees of freedom)  have the same physical meaning as the original variable(s). 
In other words, the enlarged system is obtained by attaching ancillas to the original system, see, e.g., Ref.~\cite{Budini2013}. 
In the quantum case,  such physical embedding  might be very hard (or not possible)  to construct for a given kernel  and, even when constructed, it might be very complicated to deal with.
Here we do not restrict ourselves to the physical embedding.
Instead we propose a minimal enlargement of the system, by introducing a set of \textit{non}-physical auxiliary variables, which 
leads to a new  set of equations, now local in time. \\
\indent For the $n\times n$ MK matrix in Eq.~(\ref{MKME}) we consider the following  structure
\begin{equation}\label{MK}
\mathbf{K}(t,t')=\sum_{j=1}^{k}\Gamma_je^{-\gamma_j(t-t')}\mathbf{E}_j(t)\mathbf{F}_j(t')\;,
\end{equation}
with $\Gamma_j, \gamma_j\in\mathbb{C}$ and $\mathbf{E}_j(t) = \mathbf{E}_j(t+T)$,
$\mathbf{F}_j(t) = \mathbf{F}_j(t+T) \in \mathbb{C}^{n\times n}$.
In the stationary case, $\mathbf{E}_j(t) \equiv \mathbf{E}_j$ and $\mathbf{F}_j(t') \equiv \mathbf{F}_j$,
this form relates to Erlang's method of stages \cite{Cox1955}. 
The structure given by Eq.~(\ref{MK}) allows to reproduce -- exactly or arbitrary well -- 
a large class of memory kernels, including oscillatory ones~\cite{Beylkin2005}.\\
\indent With the kernel~(\ref{MK}), the time evolution given by Eq.~(\ref{MKME}) 
for the physical system described by $\mathbf{x}$ is equivalently 
obtained by solving a time-local set of equations in which the $n$-dimensional vector $\mathbf{x}$ is coupled to an  auxiliary variable  
$\mathbf{u}$ of the dimension $p=n\times k$. The equations for the extended system read (from now on we set $t_0=0$)
\begin{equation}\label{system}
\left\{
  \begin{array}{lr}
\dot{\mathbf{x}}(t)=-\mathbf{H}(t)\mathbf{u}(t)\\
\dot{\mathbf{u}}(t)=-\mathbf{G}(t)\mathbf{x}(t)-\mathbf{A}\mathbf{u}(t)\;,
  \end{array}
\right.
\end{equation}
where $\mathbf{H}(t)=(\Gamma_1\mathbf{E}_1(t),\dots,\Gamma_k\mathbf{E}_k(t))$, $\mathbf{G}^{\rm T}(t)=(\mathbf{F}^{\rm T}_1(t),\dots,\mathbf{F}^{\rm T}_k(t))$,
and $\mathbf{A}=\rm diag(\gamma_1\mathbf{1}^{n\times n},\dots,\gamma_k\mathbf{1}^{n\times n})$.\\
\indent To assess the equivalence of Eqs.~(\ref{MKME}) and~(\ref{system}) we set $\mathbf{G}(t)\mathbf{x}(t)\equiv\mathbf{w}(t)$ and 
note that the equation for $\mathbf{u}$ reads in 
Laplace space $\mathbf{u}(\lambda)=\left[ \lambda \mathbf{1}+\mathbf{A}\right]^{-1} \mathbf{u}(0)-\left[ \lambda \mathbf{1}+\mathbf{A}\right]^{-1} \mathbf{w}(\lambda)$.
Going back to the time domain,  multiplying the resulting equation to the left by $-\mathbf{H}(t)$, and
using  the first of equations~(\ref{system}),  we end up with Eq.~(\ref{MKME}) with $\mathbf{K}(t,t')=\mathbf{H}(t) e^{-\mathbf{A}(t-t')}\mathbf{G}(t')$,
 provided that the following relation is satisfied
\begin{equation}
\mathbf{z}(t)=-\mathbf{H}(t) e^{-\mathbf{A}t}\mathbf{u}(0)\;.
\label{inhom}
\end{equation}
Thus, the requirement for the evolution of the physical part $\mathbf{x}(t)$ of the enlarged system to coincide with that given by Eq.~(\ref{MKME}),
fixes the initial condition for the auxiliary variable $\mathbf{u}(t)$.\\
\indent The system of equations~(\ref{system}) can be put into the compact form
\begin{eqnarray}\label{system-compact}
\dot{\mathbf{v}}(t)=\mathbf{M}(t)\mathbf{v}(t)\;,
\end{eqnarray}
where $\mathbf{v}=(x_1,\dots,x_n,u_1,\dots,u_{nk})^{\rm T}$ is the $(p=n+nk)$-dimensional vector describing the enlarged system, with  the matrix $\mathbf{M}$ assuming the block structure
\begin{eqnarray}\label{M}
\mathbf{M}(t)=\left(\begin{array}{lr}
\quad\mathbf{0}\quad-\mathbf{H}(t)\\
-\mathbf{G}(t)\quad\mathbf{A}
\end{array}
\right)
\end{eqnarray}
where $\mathbf{0}\in\mathbb{R}^{n\times n}$. A memory kernel of the form~(\ref{MK}), with $T$-periodic $\mathbf{E}(t)$ and $\mathbf{F}(t)$, entails the $T$-periodicity for the matrix $\mathbf{M}(t)$.
This in turn ensures that Eq.~(\ref{system-compact}) qualifies to invoke the Floquet theorem which leads to a solution of the form  $\mathbf{v}(t) = \tilde{\mathbf{S}}(t,0)e^{\mathbf{R}t}\mathbf{v}(0)$,
where $\tilde{\mathbf{S}}(t,0)$ is a $T$-periodic $p \times p$ matrix and $\mathbf{R}$  a constant $p \times p$ matrix \cite{Yakubovich1975}.
The projection of the vector $\mathbf{v}(t)$  onto the physical manifold, $\mathcal{P}_{\mathbf{x}}\mathbf{v}(t) = \mathbf{x}(t)$,   yields for the solution in the original state space,
\begin{equation}\label{STM}
\mathbf{x}(t) = \mathbf{S}(t,0)e^{\mathbf{R} t} \mathbf{v}(0)\;,
\end{equation}
with a  $n\times p$  matrix $\mathbf{S}=\mathcal{P}_{\mathbf{x}}\tilde{\mathbf{S}}$. This is the form expected from the gFT~\cite{Traversa2013}. 
In the time-local limit, i.e.,  $\mathbf{K}(t,t')=\delta(t-t')\mathbf{K}_{\rm t.l.}(t)$, the solution reduces to the standard Floquet form, with $p=n$~\cite{Yakubovich1975}.\\
\indent Equation~(\ref{system-compact}) can be solved
by constructing the Floquet propagator $\mathbf{U}_T = \mathcal{T} \exp[\int_0^T \mathbf{M}(\tau) d \tau]$,
where $\cal{T}$ denotes the time-ordering operator, and then finding its invariant, $\mathbf{U}_T\mathbf{y} = \mathbf{y}$ \cite{Yakubovich1975}.
This vector yields the asymptotic solution at  stroboscopic instants of time, i.e.,  $\mathbf{y} = \mathbf{v}^{\mathrm{a}}(sT)$, $s \in \mathbb{Z}$.
The spectral properties of the propagator can be used to characterize the relaxation time towards the asymptotic state. A conventional candidate is the spectral gap \cite{Levin2008},
$g = 1-|\lambda_m|$, with $\lambda_m$ being  the second largest (by absolute value) eigenvalue of $\mathbf{U}_T$
after $\lambda_1 = 1$. 
A straightforward diagonalization of the propagator and the use of the  obtained $\lambda_m$ as the quantifier of the relaxation speed 
is not suitable in our case. This is  because this eigenvalue  addresses the whole 
enlarged space, including the part  which is not accessible with any physically meaningful 
initial condition. To address the physical manifold only, we suggest the Arnoldi iteration method, 
starting with the initial vector  $\mathbf{v}^{\rm T}(0) = [\mathbf{x}^{\rm T}(0),\mathbf{u}^{\rm T}(0)]$, where $\mathbf{u}(0)$ satisfies Eq.~(\ref{inhom}), 
with consecutive diagonalization of the Hessenberg   matrix~\cite{Golub1996}.\\
\section{Application: Periodically driven quantum random walk}
Here we apply the method described above to the non-Markovian master equation for a continuous time quantum random walk model which yields, by construction, a completely positive and trace preserving (CPT) quantum evolution~\cite{Budini2004,Vacchini2013}. 
The model has a direct interpretation in terms of an operator generalization of a classical semi-Markov process, 
a multi-site jump process defined by a transition matrix and a waiting time distribution (WTD) $f$~\cite{Vacchini2012,Vacchini2016}. 
This classical process is itself described by a generalized master equation of the form~(\ref{MKME}) 
and turns out to be Markovian only for exponentially distributed waiting times $\tau$ between consecutive jumps, i.e., $f(\tau)=\lambda \exp(-\lambda \tau)$.\\
\indent In our application, a qubit whose density matrix $\rho(t)$ undergoes a trivial continuous background evolution (provided by the identity map $\mathbb{1}$), interrupted by the instantaneous actions of a CPT map $\mathcal{E}$ ~\cite{Budini2004,Vacchini2013}. 
These 'collisions' occur at random instances of time, with the time intervals between consecutive collisions  distributed according to a WTD  of the bi-exponential form 
\begin{eqnarray}\label{WTD}
f(\tau)=\frac{2A}{a}e^{-\gamma \tau/2}\sinh(a\tau/2)\;,
\end{eqnarray}
with $a = \sqrt{\gamma^2-4A}>0$. \\
\begin{figure}[t]
\begin{center}
\includegraphics[width=0.9\linewidth,angle=0]{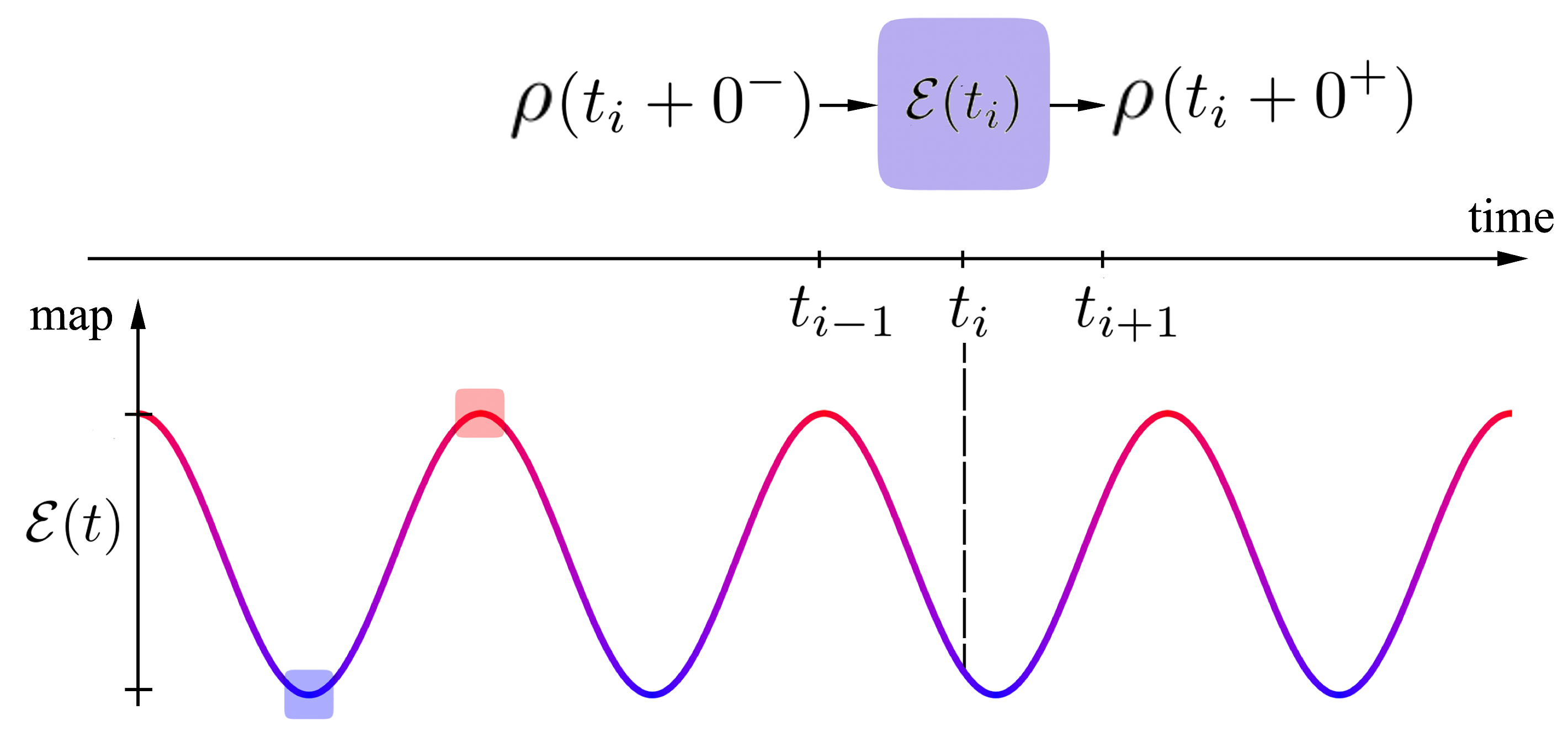}
\caption{(color online) Continuous time quantum random walk with a time-periodic map. 
A qubit undergoes repeated instantaneous actions of a time-dependent map $\mathcal{E}(t)$  at random instances of times $...t_{i-1},t_i,t_{i+1},...$.
The time interval between consecutive collisions, $ \tau_i = t_{i+1} - t_i$,
is  controlled by a  waiting time distribution $f(\tau)$. 
There is no evolution of the qubit in between  collisions. 
The  map $\mathcal{E}(t)$ is obtained as a convex combination of constant CPT maps (in this case, of two maps labelled by colored squares), with time-periodic coefficients. 
Averaging over an infinite number of realizations of the process results in the master equation~(\ref{ME}).}
\label{fig-scheme}
\end{center}
\end{figure}
\indent We generalize the original model~\cite{Budini2004,Vacchini2013}  by assuming that the collision map itself periodically evolves 
in time, $\mathcal{E}(t+T) = \mathcal{E}(t)$; see Fig.~\ref{fig-scheme}.  The time-periodic CPT $\mathcal{E}(t)$ can be constructed  as
a convex combination of $L$ CPT maps $\mathcal{E}(t)= \sum_s^L l_s(t)\mathcal{E}_s$,
where $l_s(t) \geq 0$, $\sum_s^L l_s(t) = 1$ for $\forall t \in[0,T]$, and $l_s(t+T) = l_s(t)$.
Note that, in order to get a nontrivial asymptotic state, $\rho^{\mathrm{a}}(t) \neq \id/2$, at least
one of the maps $\mathcal{E}_s$ has to be non-unital. We choose the following time-periodic map
\begin{eqnarray}\label{E}
\mathcal{E}(t)=l_1(t)\mathcal{E}_1+l_2(t)\mathcal{E}_2+l_3(t)\mathcal{E}_3\;,
\end{eqnarray}
with $l_1(t)= \sin^2 (\Omega t)$, $l_2=\sin^2(\Omega t)\cos^2(\Omega t)$, and $ l_3= \cos^4(\Omega t)$,
so that $T=\pi/\Omega$.
The map $\mathcal{E}_1$ is the non-unital amplitude damping map defined by the following action on the $2\times 2$ qubit density matrix  \cite{Nielsen2010}
\begin{eqnarray}\label{E1}
\mathcal{E}_1[\rho]&=&M_0\rho M_0^{\dag}+M_1\rho M_1^{\dag},
\end{eqnarray}
where $M_0=\begin{pmatrix}
1&0\\
0&b\\
\end{pmatrix}$ and $M_1=\begin{pmatrix}
0&d\\
0&0\\
\end{pmatrix}$
, with $b^2=1-d^2$.
The remaining two maps are  $\mathcal{E}_{2(3)}[\rho]=M_{2(3)}\rho M_{2(3)}^{\dag}$, with $M_{2(3)}=(\sigma_{y(x)}+\sigma_z)/\sqrt{2}$.\\
\indent  The density operator  of the qubit is the average over all the possible realizations of the  described process,   
\begin{eqnarray}\label{Lambda}
\rho(t)&=& g(t)\rho_0+\sum_{n=1}^{\infty}\int_0^{t}dt_{n}\dots\int_0^{t_2}dt_{1} f(t-t_n)\nonumber\\
&&\times\mathcal{E}(t_n)\dots\mathcal{E}(t_2)f(t_2-t_1)\mathcal{E}(t_1)g(t_1)\rho_0\;,\quad
\end{eqnarray}
where the function $g(t)=1-\int_0^{t}d\tau f(\tau)$ yields the probability that no collision has occurred up to time $t$.
The resulting dynamics is 
equivalently obtained as the solution of the non-Markovian master equation
\begin{eqnarray}\label{ME}
\dot{\rho}(t)=\int_0^{t}dt'\mathcal{K}(t-t')\mathcal{E}(t')\rho(t')+\mathcal{I}(t)\rho_0\;,
\end{eqnarray}
where $\mathcal{K}(t)=[\frac{d}{dt}f(t)+f(0)\delta(t)]\mathbb{1}$ and $\mathcal{I}(t)=\frac{d}{dt}g(t)\mathbb{1}=-f(t)\id$ (see the Appendix). Equation~(\ref{ME}) is one of the few known instances of a well-defined memory-kernel quantum master equation. Within the path integral formalism, evolution equations of the general form~(\ref{MKME}) are also found for driven dissipative quantum systems~\cite{Grifoni1996,Grifoni1998}. \\
\indent In order to cast Eq.~(\ref{ME}) in the form of Eq.~(\ref{MKME}), it is convenient to express the action of the  
quantum map $\mathcal{E}(t)$ on the qubit density matrix as a matrix multiplication of a four-dimensional vector $\boldsymbol{x}$ 
with a $4\times 4$ matrix $\boldsymbol{\mathcal{E}}(t)$. The vector $\mathbf{x}$ has 
components $x_i={\rm Tr}\{\rho\sigma_{i-1}\}/\sqrt{2}$ ($i=1,\dots,4$) with $\sigma_0=\id$ and $\sigma_{1,2,3}\equiv \sigma_{x,y,z}$.
In this four-dimensional representation, the time periodic matrix $\boldsymbol{\mathcal{E}}(t)$ reads
\begin{eqnarray}
\boldsymbol{\mathcal{E}}(t)=\begin{pmatrix}
1&0&0&0\\
0&bf_1(t)+f_3(t)&0&f_2(t)\\
0&0&bf_1(t)-f_2(t)&f_3(t)\\
d^2f_1(t)&f_2(t)&f_3(t)&b^2f_1(t)\\
\end{pmatrix}.\nonumber
\end{eqnarray}
The four-dimensional vector $\boldsymbol{x}(t)$ obeys Eq.~(1), with $\mathbf{z}(t) = -f(t)\boldsymbol{x}_0$,
and  a $4\times 4$ kernel matrix of the form~(\ref{MK}), with $k=2$, where $\mathbf{E}_1(t)=\mathbf{E}_2(t)=\boldsymbol{1}^{4\times 4}$ and $\mathbf{F}_1(t)=\mathbf{F}_2(t)=\boldsymbol{\mathcal{E}}(t)$, and where $\Gamma_{1,2}=\pm A\gamma_{\pm}/a$ and $\gamma_{1,2}=(\gamma\pm a)/2$, cf. Eq.~(\ref{WTD}).\\
\indent The embedding procedure with the kernel~(\ref{MK}) consisting of two terms, yields a $12$-component
vector $\mathbf{v}=(x_1,\dots,x_4,u_1,\dots,u_8)^{\text{T}}$. Its evolution is governed by the time-local Eq.~(\ref{system-compact}), with time-periodic matrix
\begin{eqnarray}\label{sys}
\mathbf{M}(t)=\begin{pmatrix}
\mathbf{0}&-\mathbf{H}\\
-\mathbf{G}(t)&\mathbf{A}\\
\end{pmatrix}.
\end{eqnarray}
Here $\mathbf{0}$ is the null ${4\times 4}$ matrix, $\mathbf{H}=\begin{pmatrix}
\Gamma_1\boldsymbol{1},&\Gamma_2\boldsymbol{1}
\end{pmatrix}$,
\begin{eqnarray}\label{SM-matrices}
\mathbf{G}(t)=\begin{pmatrix}
\boldsymbol{\mathcal{E}}(t)\\
\boldsymbol{\mathcal{E}}(t)
\end{pmatrix},\qquad\text{and}\qquad
\mathbf{A}=\begin{pmatrix}
\gamma_1\boldsymbol{1}&\mathbf{0}\\
\mathbf{0}&\gamma_2\boldsymbol{1}
\end{pmatrix},
\end{eqnarray}
where $\boldsymbol{1}\in\mathbb{R}^{4\times 4}$.
Note that the embedding procedure yields both the  transient and asymptotic dynamics for the physical variable $\mathbf{x}(t)$ provided that 
the initial condition for the auxiliary vector, determined by using Eq.~(\ref{inhom}), is
\begin{eqnarray}
\mathbf{u}(0)=-\left(\frac{x_1(0)}{\gamma_1},\dots,\frac{x_4(0)}{\gamma_1},\frac{x_1(0)}{\gamma_2},\dots,\frac{x_4(0)}{\gamma_2}\right)^{\text{T}}.\nonumber
\end{eqnarray}
\begin{figure}[t]
\begin{center}
\includegraphics[width=0.95\linewidth,angle=0]{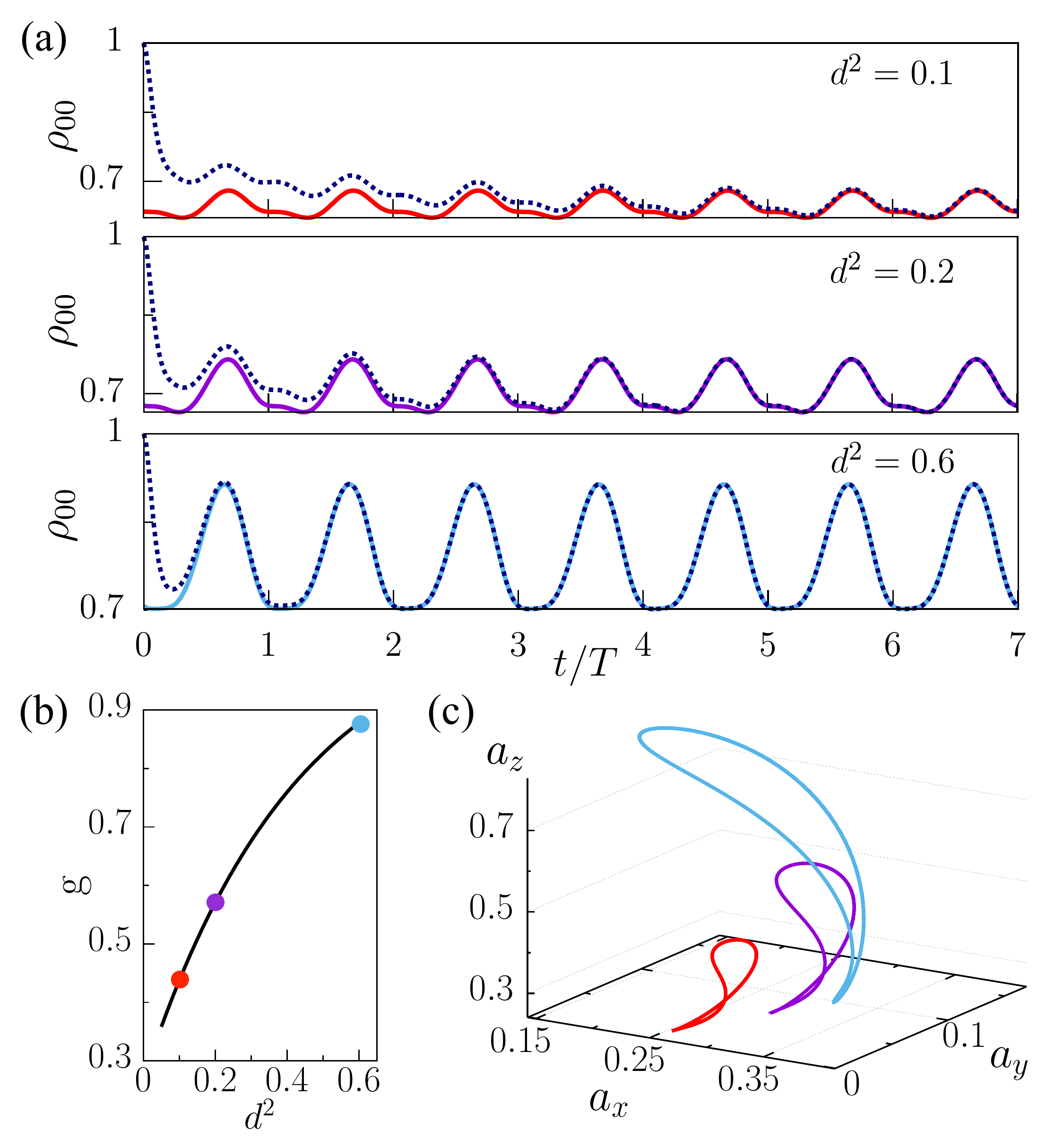}
\caption{(color online) Periodically modulated continuous time quantum random walk of a qubit. Dynamics, spectral gap, and 
limit cycle solution for three values of the parameter $d^2$ of the amplitude damping map $\mathcal{E}_1$, see Eq.~(\ref{E1}), are shown.
$(a)$  Relaxation of the state $|0\rangle$ population  obtained by numerical integration of Eq.~(\ref{MKME}) with initial condition $\rho(0)=|0\rangle\langle 0|$ (dotted lines), and asymptotic periodic states (solid lines) obtained from the Floquet propagator $\mathbf{U}_T$ of the embedded system.    
(b) Relevant spectral gap $g$ of $\mathbf{U}_T$. (c)  Limit cycle solutions for the Bloch
vector. The waiting time distribution~(\ref{WTD}) has parameters $\gamma=1$ and $A=0.24$. The modulation frequency $\Omega$ is  $0.1$.}
\label{fig2}
\end{center}
\end{figure}
\indent In Fig.~\ref{fig2}(a) we show the time evolution of $\rho_{00}(t)=\langle 0|\rho(t)|0\rangle$ obtained by the direct integration of Eq.~(\ref{MKME}), 
starting from $\rho(0)=|0\rangle\langle 0|$  (dashed lines).
After several periods, the  solutions land on the asymptotic limit cycles (solid lines) obtained from the Floquet propagator $\mathbf{U}_T$ with three Arnoldi iterations.  
The corresponding limit cycles of the Bloch vector of components $a_i={\rm Tr}\{\rho\sigma_i\}$ ($i=x,y,z$) are presented  in Fig.~\ref{fig2}(c).
As shown in Fig.~\ref{fig2}(b), the relevant spectral gap increases  with the value of  $d^2$, which in turn corresponds  to a shorter timescale of relaxation towards the
asymptotic state, see dashed lines in Fig.~\ref{fig2}(a).\\
\section{Conclusions}
We have presented a method to find the asymptotic Floquet states for a class of periodically modulated systems governed by memory-kernel master equations. 
The method has been applied to a time-periodically modulated model of piecewise dynamics of a qubit, a quantum generalization of a classical semi-Markov process \cite{Breuer2008}. 
The asymptotic Floquet states are especially interesting in this context.  In the stationary limit, the difference in non-Markovian and Markovian evolutions is
noticeable only during the relaxation towards the asymptotic stationary state~\cite{Breuer2016},  which, f.e., in the case of the qubit is a point inside (or on) Bloch sphere.
This point can be reached by following infinitely many trajectories, some of them corresponding to Markovian evolution  and some  not,
so once the relaxation is over it is impossible to decide what kind of evolution the system has undergone.
It is different  when the qubit is periodically modulated
because its asymptotic state represents a one-dimensional, time-parametrized manifold; see Fig.~\ref{fig2}(c). This manifold is specific to the Liouville superoperator $\mathcal{L}$ and it could be that some Floquet states are not attainable with Markovian $\mathcal{L}$. Outside of the quantum field,
memristors~\cite{Pershin2011} and meta-materials with memory \cite{Driscoll2009,Zheng2013} are considered now as  perspective candidates for a new generation of
nano-scale devices. They are typically modeled with equation (1); modulations can be  introduced in these systems in different ways, thus creating room for new regimes.
\section*{Acknowledgments}
 The authors gratefully acknowledge fruitful discussions with P. Talkner.   
S. D. and P. H. acknowledge the support by the Russian Science Foundation, Grant No.~15-12-20029 (S.D.)
and  by the Deutsche Forschungsgemeinschaft (DFG) via the grants
DE1889/1-1 (S.D.) and HA1517/35-1 (P.H.).

\appendix
\section{Derivation of Eq.~(\ref{ME})}
Consider the general case in which the continuous background evolution between jumps is provided by some CPT map $\mathcal{F}(t)$. 
In the application we consider the case of a continuous time quantum random walk~\cite{Budini2004} by setting $\mathcal{F}(t)\equiv\mathbb{1}$. 
The jumps are caused by the instantaneous actions of a CPT map $\mathcal{E}(t)$ at random instances of time distributed according to a waiting time distribution $f(t)$. 
The starting point for deriving the generalized master equation~(\ref{ME}) for the density matrix $\rho(t)$ in the case of modulated piecewise dynamics, meaning that the map $\mathcal{E}$ is is itself time-dependent, is the following sum over trajectories
\begin{eqnarray}\label{SME0}
\rho(t)&=&\mathcal{G}(t)\rho_0+\sum_{n=1}^{\infty}\int_0^{t}dt_{n}\dots\int_0^{t_2}dt_{1}\tilde{\mathcal{F}}(t-t_n)\mathcal{E}(t_n)\nonumber\\
&&\times\dots\mathcal{E}(t_2)\tilde{\mathcal{F}}(t_2-t_1)\mathcal{E}(t_1)\mathcal{G}(t_1)\rho_0\;,
\end{eqnarray}
where
\begin{eqnarray}\label{SME1}
\mathcal{G}(t)&=&g(t)\mathcal{F}(t)\nonumber\\
\tilde{\mathcal{F}}(t_{j+1}-t_j)
&=&f(t_{j+1}-t_j)\mathcal{F}(t_{j+1}-t_j)\;.
\end{eqnarray}
Here the function $g(t)$ gives the probability that no jump has occurred up to time $t$ and is therefore defined by $g(t)=1-\int_0^td\tau f(\tau)$.\\
\indent In order to obtain the piecewise dynamics described by Eq.~(\ref{SME0}) in the form of a master equation, we start by evaluating the series order by order in the number $n$ of \emph{jupms}, i.e., of actions of the map $\mathcal{E}$.
\begin{itemize}
\item  Zero jumps ($n=0$)
\begin{eqnarray}\label{}
\rho^{(0)}(t)=\mathcal{G}(t)\rho_0\;.
\end{eqnarray}
\item One jump ($n=1$)
\begin{eqnarray}\label{}
\rho^{(1)}(t)&=&\int_0^{t}dt_{1}\tilde{\mathcal{F}}(t-t_1)\mathcal{E}(t_1)\mathcal{G}(t_1)\rho_0\nonumber\\
&=&\int_0^{t}dt_{1}\tilde{\mathcal{F}}(t-t_1)\mathcal{E}(t_1)\rho^{(0)}(t_1)\;.
\end{eqnarray}
\item Two jumps ($n=2$)
\begin{eqnarray}\label{}
\rho^{(2)}(t)&=&\int_0^{t}dt_{2}\int_0^{t_2}dt_{1}\tilde{\mathcal{F}}(t-t_2)\mathcal{E}(t_2)\nonumber\\
&&\quad\times\tilde{\mathcal{F}}(t_2-t_1)\mathcal{E}(t_1)\mathcal{G}(t_1)\rho_0\nonumber\\
&=&\int_0^{t}dt_{2}\tilde{\mathcal{F}}(t-t_2)\mathcal{E}(t_2)\nonumber\\
&&\quad\times\int_0^{t_2}dt_{1}\tilde{\mathcal{F}}(t_2-t_1)\mathcal{E}(t_1)\rho^{(0)}(t_1)\nonumber\\
&=&\int_0^{t}dt_{2}\tilde{\mathcal{F}}(t-t_2)\mathcal{E}(t_2)\rho^{(1)}(t_2)\;,
\end{eqnarray}
\end{itemize}
and so on. We find the recursive relation
\begin{eqnarray}\label{}
\rho^{(n)}(t)&=&\int_0^t dt'\tilde{\mathcal{F}}(t-t')\mathcal{E}(t')\rho^{(n-1)}(t'),\nonumber \qquad(n\geq 1)\nonumber\\
\rho^{(0)}(t)&=&\mathcal{G}(t)\rho_0\;.
\end{eqnarray}
Summing the series we get
\begin{eqnarray}\label{Srho-t-2}
\rho(t)&=&\mathcal{G}(t)\rho_0+\int_0^t dt'\tilde{\mathcal{F}}(t-t')\mathcal{E}(t')\sum_{n=1}^{\infty}\rho^{(n-1)}(t')\nonumber\\
&=&\mathcal{G}(t)\rho_0+\int_0^t dt'\tilde{\mathcal{F}}(t-t')\mathcal{E}(t')\rho(t')\;.
\end{eqnarray}
Finally, taking the time derivative of Eq.~(\ref{Srho-t-2}) we obtain
\begin{eqnarray}\label{SME}
\dot{\rho}(t)&=&\frac{d}{d t}\mathcal{G}(t)\rho_0+\int_0^{t}dt'\frac{\partial}{\partial t}\tilde{\mathcal{F}}(t-t')\mathcal{E}(t')\rho(t')\nonumber\\
&&+\tilde{\mathcal{F}}(0)\mathcal{E}(t)\rho(t)\nonumber\\
&=&\int_0^{t}dt'\mathcal{K}(t-t')\mathcal{E}(t')\rho(t')+\mathcal{I}(t)\rho_0\;,
\end{eqnarray}
where
\begin{eqnarray}\label{SME-2}
\mathcal{K}(t)&=&\frac{d}{d t}\tilde{\mathcal{F}}(t)+\tilde{\mathcal{F}}(0)\delta(t)\nonumber\\
\mathcal{I}(t)&=&\frac{d}{d t}\mathcal{G}(t)\;.
\end{eqnarray}
In the static case $\mathcal{E}(t)\equiv\mathcal{E}$, Eq.~(\ref{SME})  coincides with Eq. (7) of Ref.~\cite{Vacchini2013}. If instead we set  $\mathcal{F}(t)\equiv\mathbb{1}$ the case considered in the application is recovered.


\begin{thebibliography}{52}%
\makeatletter
\providecommand \@ifxundefined [1]{%
 \@ifx{#1\undefined}
}%
\providecommand \@ifnum [1]{%
 \ifnum #1\expandafter \@firstoftwo
 \else \expandafter \@secondoftwo
 \fi
}%
\providecommand \@ifx [1]{%
 \ifx #1\expandafter \@firstoftwo
 \else \expandafter \@secondoftwo
 \fi
}%
\providecommand \natexlab [1]{#1}%
\providecommand \enquote  [1]{``#1''}%
\providecommand \bibnamefont  [1]{#1}%
\providecommand \bibfnamefont [1]{#1}%
\providecommand \citenamefont [1]{#1}%
\providecommand \href@noop [0]{\@secondoftwo}%
\providecommand \href [0]{\begingroup \@sanitize@url \@href}%
\providecommand \@href[1]{\@@startlink{#1}\@@href}%
\providecommand \@@href[1]{\endgroup#1\@@endlink}%
\providecommand \@sanitize@url [0]{\catcode `\\12\catcode `\$12\catcode
  `\&12\catcode `\#12\catcode `\^12\catcode `\_12\catcode `\%12\relax}%
\providecommand \@@startlink[1]{}%
\providecommand \@@endlink[0]{}%
\providecommand \url  [0]{\begingroup\@sanitize@url \@url }%
\providecommand \@url [1]{\endgroup\@href {#1}{\urlprefix }}%
\providecommand \urlprefix  [0]{URL }%
\providecommand \Eprint [0]{\href }%
\providecommand \doibase [0]{http://dx.doi.org/}%
\providecommand \selectlanguage [0]{\@gobble}%
\providecommand \bibinfo  [0]{\@secondoftwo}%
\providecommand \bibfield  [0]{\@secondoftwo}%
\providecommand \translation [1]{[#1]}%
\providecommand \BibitemOpen [0]{}%
\providecommand \bibitemStop [0]{}%
\providecommand \bibitemNoStop [0]{.\EOS\space}%
\providecommand \EOS [0]{\spacefactor3000\relax}%
\providecommand \BibitemShut  [1]{\csname bibitem#1\endcsname}%
\let\auto@bib@innerbib\@empty
\bibitem{Kadanoff1993}
  L. P. Kadanoff, \textit{From Order to Chaos} (World Scientific, 1993).
\bibitem [{\citenamefont {Petrov}\ \emph {et~al.}(1997)\citenamefont {Petrov},
  \citenamefont {Ouyang},\ and\ \citenamefont {Swinney}}]{Petrov1997}%
  \BibitemOpen
  \bibfield  {author} {\bibinfo {author} {\bibfnamefont {V.}~\bibnamefont
  {Petrov}}, \bibinfo {author} {\bibfnamefont {Q.}~\bibnamefont {Ouyang}}, \
  and\ \bibinfo {author} {\bibfnamefont {H.~L.}\ \bibnamefont {Swinney}},\
  }\href@noop {} {\bibfield  {journal} {\bibinfo  {journal} {Nature}\ }\textbf
  {\bibinfo {volume} {388}},\ \bibinfo {pages} {655} (\bibinfo {year}
  {1997})}\BibitemShut {NoStop}%
\bibitem [{\citenamefont {Herrmann}(2001)}]{Herrmann2001}%
  \BibitemOpen
  \bibfield  {author} {\bibinfo {author} {\bibfnamefont {C.~S.}\ \bibnamefont
  {Herrmann}},\ }\href {\doibase 10.1007/s002210100682} {\bibfield  {journal}
  {\bibinfo  {journal} {Exp. Brain Res.}\ }\textbf {\bibinfo {volume} {137}},\
  \bibinfo {pages} {346} (\bibinfo {year} {2001})}\BibitemShut {NoStop}%
\bibitem [{\citenamefont {Shirley}(1965)}]{Shirley1965}%
  \BibitemOpen
  \bibfield  {author} {\bibinfo {author} {\bibfnamefont {J.~H.}\ \bibnamefont
  {Shirley}},\ }\href {http://link.aps.org/doi/10.1103/PhysRev.138.B979}
  {\bibfield  {journal} {\bibinfo  {journal} {Phys. Rev.}\ }\textbf {\bibinfo
  {volume} {138}},\ \bibinfo {pages} {B979} (\bibinfo {year}
  {1965})}\BibitemShut {NoStop}%
\bibitem [{\citenamefont {Sambe}(1973)}]{Sambe1973}%
  \BibitemOpen
  \bibfield  {author} {\bibinfo {author} {\bibfnamefont {H.}~\bibnamefont
  {Sambe}},\ }\href {http://link.aps.org/doi/10.1103/PhysRevA.7.2203}
  {\bibfield  {journal} {\bibinfo  {journal} {Phys. Rev. A}\ }\textbf {\bibinfo
  {volume} {7}},\ \bibinfo {pages} {2203} (\bibinfo {year} {1973})}\BibitemShut
  {NoStop}%
\bibitem [{\citenamefont {Grifoni}\ and\ \citenamefont
  {H{\"a}nggi}(1998)}]{Grifoni1998}%
  \BibitemOpen
  \bibfield  {author} {\bibinfo {author} {\bibfnamefont {M.}~\bibnamefont
  {Grifoni}}\ and\ \bibinfo {author} {\bibfnamefont {P.}~\bibnamefont
  {H{\"a}nggi}},\ }\href {\doibase 10.1016/S0370-1573(98)00022-2} {\bibfield
  {journal} {\bibinfo  {journal} {Phys. Rep.}\ }\textbf {\bibinfo {volume}
  {304}},\ \bibinfo {pages} {229} (\bibinfo {year} {1998})}\BibitemShut
  {NoStop}%
\bibitem [{\citenamefont {Bukov}\ \emph {et~al.}(2015)\citenamefont {Bukov},
  \citenamefont {D'Alessio},\ and\ \citenamefont {Polkovnikov}}]{Bukov2015}%
  \BibitemOpen
  \bibfield  {author} {\bibinfo {author} {\bibfnamefont {M.}~\bibnamefont
  {Bukov}}, \bibinfo {author} {\bibfnamefont {L.}~\bibnamefont {D'Alessio}}, \
  and\ \bibinfo {author} {\bibfnamefont {A.}~\bibnamefont {Polkovnikov}},\
  }\href {\doibase 10.1080/00018732.2015.1055918} {\bibfield  {journal}
  {\bibinfo  {journal} {Adv. Phys.}\ }\textbf {\bibinfo {volume} {64}},\
  \bibinfo {pages} {139} (\bibinfo {year} {2015})}\BibitemShut {NoStop}%
\bibitem [{\citenamefont {Eckardt}\ and\ \citenamefont
  {Anisimovas}(2015)}]{Eckardt2015}%
  \BibitemOpen
  \bibfield  {author} {\bibinfo {author} {\bibfnamefont {A.}~\bibnamefont
  {Eckardt}}\ and\ \bibinfo {author} {\bibfnamefont {E.}~\bibnamefont
  {Anisimovas}},\ }\href {http://stacks.iop.org/1367-2630/17/i=9/a=093039}
  {\bibfield  {journal} {\bibinfo  {journal} {New J. Phys.}\ }\textbf {\bibinfo
  {volume} {17}},\ \bibinfo {pages} {093039} (\bibinfo {year}
  {2015})}\BibitemShut {NoStop}%
\bibitem [{\citenamefont {Lindner}\ \emph {et~al.}(2011)\citenamefont
  {Lindner}, \citenamefont {Refael},\ and\ \citenamefont
  {Galitski}}]{Lindner2011}%
  \BibitemOpen
  \bibfield  {author} {\bibinfo {author} {\bibfnamefont {N.~H.}\ \bibnamefont
  {Lindner}}, \bibinfo {author} {\bibfnamefont {G.}~\bibnamefont {Refael}}, \
  and\ \bibinfo {author} {\bibfnamefont {V.}~\bibnamefont {Galitski}},\ }\href
  {http://dx.doi.org/10.1038/nphys1926} {\bibfield  {journal} {\bibinfo
  {journal} {Nat. Phys.}\ }\textbf {\bibinfo {volume} {7}},\ \bibinfo {pages}
  {490} (\bibinfo {year} {2011})}\BibitemShut {NoStop}%
\bibitem [{\citenamefont {Liu}\ \emph {et~al.}(2013)\citenamefont {Liu},
  \citenamefont {Levchenko},\ and\ \citenamefont {Baranger}}]{Liu2013}%
  \BibitemOpen
  \bibfield  {author} {\bibinfo {author} {\bibfnamefont {D.~E.}\ \bibnamefont
  {Liu}}, \bibinfo {author} {\bibfnamefont {A.}~\bibnamefont {Levchenko}}, \
  and\ \bibinfo {author} {\bibfnamefont {H.~U.}\ \bibnamefont {Baranger}},\
  }\href {https://link.aps.org/doi/10.1103/PhysRevLett.111.047002} {\bibfield
  {journal} {\bibinfo  {journal} {Phys. Rev. Lett.}\ }\textbf {\bibinfo
  {volume} {111}},\ \bibinfo {pages} {047002} (\bibinfo {year}
  {2013})}\BibitemShut {NoStop}%
\bibitem [{\citenamefont {Struck}\ \emph {et~al.}(2011)\citenamefont {Struck},
  \citenamefont {{\"O}lschl{\"a}ger}, \citenamefont {{Le Targat}},
  \citenamefont {Soltan-Panahi}, \citenamefont {Eckardt}, \citenamefont
  {Lewenstein}, \citenamefont {Windpassinger},\ and\ \citenamefont
  {Sengstock}}]{Struck2011}%
  \BibitemOpen
  \bibfield  {author} {\bibinfo {author} {\bibfnamefont {J.}~\bibnamefont
  {Struck}}, \bibinfo {author} {\bibfnamefont {C.}~\bibnamefont
  {{\"O}lschl{\"a}ger}}, \bibinfo {author} {\bibfnamefont {R.}~\bibnamefont
  {{Le Targat}}}, \bibinfo {author} {\bibfnamefont {P.}~\bibnamefont
  {Soltan-Panahi}}, \bibinfo {author} {\bibfnamefont {A.}~\bibnamefont
  {Eckardt}}, \bibinfo {author} {\bibfnamefont {M.}~\bibnamefont {Lewenstein}},
  \bibinfo {author} {\bibfnamefont {P.}~\bibnamefont {Windpassinger}}, \ and\
  \bibinfo {author} {\bibfnamefont {K.}~\bibnamefont {Sengstock}},\ }\href
  {http://science.sciencemag.org/content/333/6045/996.abstract} {\bibfield
  {journal} {\bibinfo  {journal} {Science}\ }\textbf {\bibinfo {volume}
  {333}},\ \bibinfo {pages} {996} (\bibinfo {year} {2011})}\BibitemShut
  {NoStop}%
\bibitem [{\citenamefont {Goldman}\ and\ \citenamefont
  {Dalibard}(2014)}]{Goldman2014}%
  \BibitemOpen
  \bibfield  {author} {\bibinfo {author} {\bibfnamefont {N.}~\bibnamefont
  {Goldman}}\ and\ \bibinfo {author} {\bibfnamefont {J.}~\bibnamefont
  {Dalibard}},\ }\href {http://link.aps.org/doi/10.1103/PhysRevX.4.031027}
  {\bibfield  {journal} {\bibinfo  {journal} {Phys. Rev. X}\ }\textbf {\bibinfo
  {volume} {4}},\ \bibinfo {pages} {031027} (\bibinfo {year}
  {2014})}\BibitemShut {NoStop}%
\bibitem [{\citenamefont {Else}\ \emph {et~al.}(2016)\citenamefont {Else},
  \citenamefont {Bauer},\ and\ \citenamefont {Nayak}}]{Else2016}%
  \BibitemOpen
  \bibfield  {author} {\bibinfo {author} {\bibfnamefont {D.~V.}\ \bibnamefont
  {Else}}, \bibinfo {author} {\bibfnamefont {B.}~\bibnamefont {Bauer}}, \ and\
  \bibinfo {author} {\bibfnamefont {C.}~\bibnamefont {Nayak}},\ }\href
  {https://link.aps.org/doi/10.1103/PhysRevLett.117.090402} {\bibfield
  {journal} {\bibinfo  {journal} {Phys. Rev. Lett.}\ }\textbf {\bibinfo
  {volume} {117}},\ \bibinfo {pages} {090402} (\bibinfo {year}
  {2016})}\BibitemShut {NoStop}%
\bibitem [{\citenamefont {Zhang}\ \emph {et~al.}(2017)\citenamefont {Zhang}
  \emph {et~al.}}]{Zhang2017}%
  \BibitemOpen
  \bibfield  {author} {\bibinfo {author} {\bibfnamefont {J.}~\bibnamefont
  {Zhang}} \emph {et~al.},\ }\href {http://dx.doi.org/10.1038/nature21413}
  {\bibfield  {journal} {\bibinfo  {journal} {Nature}\ }\textbf {\bibinfo
  {volume} {543}},\ \bibinfo {pages} {217} (\bibinfo {year}
  {2017})}\BibitemShut {NoStop}%
\bibitem [{\citenamefont {Choi}\ \emph {et~al.}(2017)\citenamefont {Choi} \emph
  {et~al.}}]{Choi2017}%
  \BibitemOpen
  \bibfield  {author} {\bibinfo {author} {\bibfnamefont {S.}~\bibnamefont
  {Choi}} \emph {et~al.},\ }\href {http://dx.doi.org/10.1038/nature21426}
  {\bibfield  {journal} {\bibinfo  {journal} {Nature}\ }\textbf {\bibinfo
  {volume} {543}},\ \bibinfo {pages} {221} (\bibinfo {year}
  {2017})}\BibitemShut {NoStop}%
\bibitem [{\citenamefont {Floquet}(1883)}]{Floquet1883}%
  \BibitemOpen
  \bibfield  {author} {\bibinfo {author} {\bibfnamefont {G.}~\bibnamefont
  {Floquet}},\ }\href@noop {} {\bibfield  {journal} {\bibinfo  {journal} {Ann.
  Sci. Ec. Norm. Sup.}\ }\textbf {\bibinfo {volume} {12}},\ \bibinfo {pages}
  {47} (\bibinfo {year} {1883})}\BibitemShut {NoStop}%
\bibitem [{\citenamefont {Yakubovich}\ and\ \citenamefont
  {Starzhinskii}(1975)}]{Yakubovich1975}%
  \BibitemOpen
  \bibfield  {author} {\bibinfo {author} {\bibfnamefont {V.~A.}\ \bibnamefont
  {Yakubovich}}\ and\ \bibinfo {author} {\bibfnamefont {V.~M.}\ \bibnamefont
  {Starzhinskii}},\ }\href@noop {} {\emph {\bibinfo {title} {{Linear
  Differential Equations with Periodic Coeffcients}}}}\ (\bibinfo  {publisher}
  {Wiley, New York},\ \bibinfo {year} {1975})\BibitemShut {NoStop}%
\bibitem [{\citenamefont {Jung}(1993)}]{Jung1993}%
  \BibitemOpen
  \bibfield  {author} {\bibinfo {author} {\bibfnamefont {P.}~\bibnamefont
  {Jung}},\ }\href {\doibase 10.1016/0370-1573(93)90022-6} {\bibfield
  {journal} {\bibinfo  {journal} {Phys. Rep.}\ }\textbf {\bibinfo {volume}
  {234}},\ \bibinfo {pages} {175} (\bibinfo {year} {1993})}\BibitemShut
  {NoStop}%
\bibitem [{\citenamefont {Gammaitoni}\ \emph {et~al.}(1998)\citenamefont
  {Gammaitoni}, \citenamefont {H{\"a}nggi}, \citenamefont {Jung},\ and\
  \citenamefont {Marchesoni}}]{Gammaitoni1998}%
  \BibitemOpen
  \bibfield  {author} {\bibinfo {author} {\bibfnamefont {L.}~\bibnamefont
  {Gammaitoni}}, \bibinfo {author} {\bibfnamefont {P.}~\bibnamefont
  {H{\"a}nggi}}, \bibinfo {author} {\bibfnamefont {P.}~\bibnamefont {Jung}}, \
  and\ \bibinfo {author} {\bibfnamefont {F.}~\bibnamefont {Marchesoni}},\
  }\href {http://link.aps.org/doi/10.1103/RevModPhys.70.223} {\bibfield
  {journal} {\bibinfo  {journal} {Rev. Mod. Phys.}\ }\textbf {\bibinfo {volume}
  {70}},\ \bibinfo {pages} {223} (\bibinfo {year} {1998})}\BibitemShut
  {NoStop}%
\bibitem [{\citenamefont {Hartmann}\ \emph {et~al.}(2017)\citenamefont
  {Hartmann}, \citenamefont {Poletti}, \citenamefont {Ivanchenko},
  \citenamefont {Denisov},\ and\ \citenamefont {H\"anggi}}]{Hartmann2017}%
  \BibitemOpen
  \bibfield  {author} {\bibinfo {author} {\bibfnamefont {M.}~\bibnamefont
  {Hartmann}}, \bibinfo {author} {\bibfnamefont {D.}~\bibnamefont {Poletti}},
  \bibinfo {author} {\bibfnamefont {M.}~\bibnamefont {Ivanchenko}}, \bibinfo
  {author} {\bibfnamefont {S.}~\bibnamefont {Denisov}}, \ and\ \bibinfo
  {author} {\bibfnamefont {P.}~\bibnamefont {H\"anggi}},\ }\href
  {http://stacks.iop.org/1367-2630/19/i=8/a=083011} {\bibfield  {journal}
  {\bibinfo  {journal} {New J. Phys.}\ }\textbf {\bibinfo {volume} {19}},\
  \bibinfo {pages} {083011} (\bibinfo {year} {2017})}\BibitemShut {NoStop}%
\bibitem [{\citenamefont {Vorberg}\ \emph {et~al.}(2013)\citenamefont
  {Vorberg}, \citenamefont {Wustmann}, \citenamefont {Ketzmerick},\ and\
  \citenamefont {Eckardt}}]{Vorberg2013}%
  \BibitemOpen
  \bibfield  {author} {\bibinfo {author} {\bibfnamefont {D.}~\bibnamefont
  {Vorberg}}, \bibinfo {author} {\bibfnamefont {W.}~\bibnamefont {Wustmann}},
  \bibinfo {author} {\bibfnamefont {R.}~\bibnamefont {Ketzmerick}}, \ and\
  \bibinfo {author} {\bibfnamefont {A.}~\bibnamefont {Eckardt}},\ }\href
  {https://link.aps.org/doi/10.1103/PhysRevLett.111.240405} {\bibfield
  {journal} {\bibinfo  {journal} {Phys. Rev. Lett.}\ }\textbf {\bibinfo
  {volume} {111}},\ \bibinfo {pages} {240405} (\bibinfo {year}
  {2013})}\BibitemShut {NoStop}%
\bibitem [{\citenamefont {Lazarides}\ and\ \citenamefont
  {Moessner}(2017)}]{Lazarides2017}%
  \BibitemOpen
  \bibfield  {author} {\bibinfo {author} {\bibfnamefont {A.}~\bibnamefont
  {Lazarides}}\ and\ \bibinfo {author} {\bibfnamefont {R.}~\bibnamefont
  {Moessner}},\ }\href@noop {} {\bibfield  {journal} {\bibinfo  {journal}
  {arXiv:1703.02547}\ } (\bibinfo {year} {2017})}\BibitemShut {NoStop}%
\bibitem [{\citenamefont {Sommer}\ and\ \citenamefont
  {Simon}(2016)}]{Sommer2016}%
  \BibitemOpen
  \bibfield  {author} {\bibinfo {author} {\bibfnamefont {A.}~\bibnamefont
  {Sommer}}\ and\ \bibinfo {author} {\bibfnamefont {J.}~\bibnamefont {Simon}},\
  }\href {http://stacks.iop.org/1367-2630/18/i=3/a=035008} {\bibfield
  {journal} {\bibinfo  {journal} {New J. Phys.}\ }\textbf {\bibinfo {volume}
  {18}},\ \bibinfo {pages} {035008} (\bibinfo {year} {2016})}\BibitemShut
  {NoStop}%
\bibitem [{\citenamefont {Meinert}\ \emph {et~al.}(2016)\citenamefont
  {Meinert}, \citenamefont {Mark}, \citenamefont {Lauber}, \citenamefont
  {Daley},\ and\ \citenamefont {N{\"a}gerl}}]{Meinert2016}%
  \BibitemOpen
  \bibfield  {author} {\bibinfo {author} {\bibfnamefont {F.}~\bibnamefont
  {Meinert}}, \bibinfo {author} {\bibfnamefont {M.~J.}\ \bibnamefont {Mark}},
  \bibinfo {author} {\bibfnamefont {K.}~\bibnamefont {Lauber}}, \bibinfo
  {author} {\bibfnamefont {A.~J.}\ \bibnamefont {Daley}}, \ and\ \bibinfo
  {author} {\bibfnamefont {H.~C.}\ \bibnamefont {N{\"a}gerl}},\ }\href
  {http://link.aps.org/doi/10.1103/PhysRevLett.116.205301} {\bibfield
  {journal} {\bibinfo  {journal} {Phys. Rev. Lett.}\ }\textbf {\bibinfo
  {volume} {116}},\ \bibinfo {pages} {205301} (\bibinfo {year}
  {2016})}\BibitemShut {NoStop}%
\bibitem [{\citenamefont {Eckardt}(2017)}]{Eckardt2017}%
  \BibitemOpen
  \bibfield  {author} {\bibinfo {author} {\bibfnamefont {A.}~\bibnamefont
  {Eckardt}},\ }\href {https://link.aps.org/doi/10.1103/RevModPhys.89.011004}
  {\bibfield  {journal} {\bibinfo  {journal} {Rev. Mod. Phys.}\ }\textbf
  {\bibinfo {volume} {89}},\ \bibinfo {pages} {011004} (\bibinfo {year}
  {2017})}\BibitemShut {NoStop}%
\bibitem [{\citenamefont {Kohler}\ \emph {et~al.}(2005)\citenamefont {Kohler},
  \citenamefont {Lehmann},\ and\ \citenamefont {H{\"a}nggi}}]{Kohler2005}%
  \BibitemOpen
  \bibfield  {author} {\bibinfo {author} {\bibfnamefont {S.}~\bibnamefont
  {Kohler}}, \bibinfo {author} {\bibfnamefont {J.}~\bibnamefont {Lehmann}}, \
  and\ \bibinfo {author} {\bibfnamefont {P.}~\bibnamefont {H{\"a}nggi}},\
  }\href {\doibase 10.1016/j.physrep.2004.11.002} {\bibfield  {journal}
  {\bibinfo  {journal} {Phys. Rep.}\ }\textbf {\bibinfo {volume} {406}},\
  \bibinfo {pages} {379} (\bibinfo {year} {2005})}\BibitemShut {NoStop}%
\bibitem [{\citenamefont {Alicki}\ and\ \citenamefont
  {Lendi}(2007)}]{Alicki2007}%
  \BibitemOpen
  \bibfield  {author} {\bibinfo {author} {\bibfnamefont {R.}~\bibnamefont
  {Alicki}}\ and\ \bibinfo {author} {\bibfnamefont {K.}~\bibnamefont {Lendi}},\
  }\href@noop {} {\emph {\bibinfo {title} {{Quantum Dynamical Semigroups and
  Applications}}}},\ \bibinfo {series} {{Lecture Notes in Physics}}, Vol.\
  \bibinfo {volume} {717}\ (\bibinfo  {publisher} {Springer Berlin
  Heidelberg},\ \bibinfo {year} {2007})\BibitemShut {NoStop}%
\bibitem [{\citenamefont {Bastidas}\ \emph {et~al.}(2017)\citenamefont
  {Bastidas}, \citenamefont {Kyaw}, \citenamefont {Tangpanitanon},
  \citenamefont {Romero}, \citenamefont {Kwek},\ and\ \citenamefont
  {Angelakis}}]{Bastidas2017}%
  \BibitemOpen
  \bibfield  {author} {\bibinfo {author} {\bibfnamefont {V.~M.}\ \bibnamefont
  {Bastidas}}, \bibinfo {author} {\bibfnamefont {T.~H.}\ \bibnamefont {Kyaw}},
  \bibinfo {author} {\bibfnamefont {J.}~\bibnamefont {Tangpanitanon}}, \bibinfo
  {author} {\bibfnamefont {G.}~\bibnamefont {Romero}}, \bibinfo {author}
  {\bibfnamefont {L.-C.}\ \bibnamefont {Kwek}}, \ and\ \bibinfo {author}
  {\bibfnamefont {D.~G.}\ \bibnamefont {Angelakis}},\ }\href@noop {} {\bibfield
   {journal} {\bibinfo  {journal} {arXiv:1707.04423v1}\ } (\bibinfo {year}
  {2017})}\BibitemShut {NoStop}%
\bibitem [{\citenamefont {Breuer}\ \emph {et~al.}(2016)\citenamefont {Breuer},
  \citenamefont {Laine}, \citenamefont {Piilo},\ and\ \citenamefont
  {Vacchini}}]{Breuer2016}%
  \BibitemOpen
  \bibfield  {author} {\bibinfo {author} {\bibfnamefont {H.-P.}\ \bibnamefont
  {Breuer}}, \bibinfo {author} {\bibfnamefont {E.-M.}\ \bibnamefont {Laine}},
  \bibinfo {author} {\bibfnamefont {J.}~\bibnamefont {Piilo}}, \ and\ \bibinfo
  {author} {\bibfnamefont {B.}~\bibnamefont {Vacchini}},\ }\href {\doibase
  10.1103/RevModPhys.88.021002} {\bibfield  {journal} {\bibinfo  {journal}
  {Rev. Mod. Phys.}\ }\textbf {\bibinfo {volume} {88}},\ \bibinfo {pages}
  {021002} (\bibinfo {year} {2016})}\BibitemShut {NoStop}%
\bibitem{Liu2011}
  Bi-Heng Liu, Li Li, Yun-Feng Huang, Chuan-Feng Li, Guang-Can Guo, E.-M.Laine, H.-P. Breuer, J. Piilo,
  Nature Phys. \textbf{7}, 931 (2011). 
\bibitem [{\citenamefont {Traversa}\ \emph {et~al.}(2013)\citenamefont
  {Traversa}, \citenamefont {{Di Ventra}},\ and\ \citenamefont
  {Bonani}}]{Traversa2013}%
  \BibitemOpen
  \bibfield  {author} {\bibinfo {author} {\bibfnamefont {F.~L.}\ \bibnamefont
  {Traversa}}, \bibinfo {author} {\bibfnamefont {M.}~\bibnamefont {{Di
  Ventra}}}, \ and\ \bibinfo {author} {\bibfnamefont {F.}~\bibnamefont
  {Bonani}},\ }\href {http://link.aps.org/doi/10.1103/PhysRevLett.110.170602}
  {\bibfield  {journal} {\bibinfo  {journal} {Phys. Rev. Lett.}\ }\textbf
  {\bibinfo {volume} {110}},\ \bibinfo {pages} {170602} (\bibinfo {year}
  {2013})}\BibitemShut {NoStop}%
\bibitem [{\citenamefont {Budini}(2004)}]{Budini2004}%
  \BibitemOpen
  \bibfield  {author} {\bibinfo {author} {\bibfnamefont {A.~A.}~\bibnamefont
  {Budini}},\ }\href {http://link.aps.org/doi/10.1103/PhysRevA.69.042107}
  {\bibfield  {journal} {\bibinfo  {journal} {Phys. Rev. A}\ }\textbf {\bibinfo
  {volume} {69}},\ \bibinfo {pages} {042107} (\bibinfo {year}
  {2004})}\BibitemShut {NoStop}%
\bibitem [{Note1()}]{Note1}%
  \BibitemOpen
  \bibinfo {note} {For $\protect \qopname \relax m{lim}_{t \rightarrow \infty
  }\protect \mathbf {z}(t+T) = \protect \mathbf {z}(t)\not =\protect \mathbf
  {0}$, the method has to be complemented with a non-homogeneous extension of
  Floquet theory \cite {Yakubovich1975}.}\BibitemShut {Stop}%
\bibitem [{\citenamefont {Vacchini}(2013)}]{Vacchini2013}%
  \BibitemOpen
  \bibfield  {author} {\bibinfo {author} {\bibfnamefont {B.}~\bibnamefont
  {Vacchini}},\ }\href {http://link.aps.org/doi/10.1103/PhysRevA.87.030101}
  {\bibfield  {journal} {\bibinfo  {journal} {Phys. Rev. A}\ }\textbf {\bibinfo
  {volume} {87}},\ \bibinfo {pages} {030101} (\bibinfo {year}
  {2013})}\BibitemShut {NoStop}%
\bibitem [{\citenamefont {Vacchini}(2016)}]{Vacchini2016}%
  \BibitemOpen
  \bibfield  {author} {\bibinfo {author} {\bibfnamefont {B.}~\bibnamefont
  {Vacchini}},\ }\href {\doibase 10.1103/PhysRevLett.117.230401} {\bibfield
  {journal} {\bibinfo  {journal} {Phys. Rev. Lett.}\ }\textbf {\bibinfo
  {volume} {117}},\ \bibinfo {pages} {230401} (\bibinfo {year}
  {2016})}\BibitemShut {NoStop}%
\bibitem{note1}  Being preconditioned by the existence of the unique $T$-periodic asymptotic solution 
$\mathbf{x}^{\mathrm{a}}(t+T) = \mathbf{x}^{\mathrm{a}}(t)$, 
Eq.~(\ref{MKME}) can be transformed into the finite-length  memory  form used in Ref.~\cite{Traversa2013}, with $r = T$, by folding the kernel into the time interval $[t,t-T]$,
$\mathbf{\tilde{K}}(t,t') = \sum_{s=0}^{\infty} \mathbf{K}(t, t' - sT)$, and setting $\mathbf{z}(t) =  \mathbf{0}$.    
\bibitem [{\citenamefont {Grabert}\ \emph {et~al.}(1977)\citenamefont
  {Grabert}, \citenamefont {Talkner},\ and\ \citenamefont
  {H{\"a}nggi}}]{Grabert1977}%
  \BibitemOpen
  \bibfield  {author} {\bibinfo {author} {\bibfnamefont {H.}~\bibnamefont
  {Grabert}}, \bibinfo {author} {\bibfnamefont {P.}~\bibnamefont {Talkner}}, \
  and\ \bibinfo {author} {\bibfnamefont {P.}~\bibnamefont {H{\"a}nggi}},\
  }\href {\doibase 10.1007/BF01570749} {\bibfield  {journal} {\bibinfo
  {journal} {Z. Physik B}\ }\textbf {\bibinfo {volume} {26}},\ \bibinfo {pages}
  {389} (\bibinfo {year} {1977})}\BibitemShut {NoStop}%
\bibitem [{\citenamefont {Grabert}\ \emph {et~al.}(1978)\citenamefont
  {Grabert}, \citenamefont {Talkner},\ and\ \citenamefont
  {H{\"a}nggi}}]{Grabert1978}%
  \BibitemOpen
  \bibfield  {author} {\bibinfo {author} {\bibfnamefont {H.}~\bibnamefont
  {Grabert}}, \bibinfo {author} {\bibfnamefont {P.}~\bibnamefont {Talkner}}, \
  and\ \bibinfo {author} {\bibfnamefont {P.}~\bibnamefont {H{\"a}nggi}},\
  }\href {\doibase 10.1007/BF01321192} {\bibfield  {journal} {\bibinfo
  {journal} {Z. Physik B}\ }\textbf {\bibinfo {volume} {29}},\ \bibinfo {pages}
  {273} (\bibinfo {year} {1978})}\BibitemShut {NoStop}%
\bibitem [{\citenamefont {Grabert}\ \emph {et~al.}(1980)\citenamefont
  {Grabert}, \citenamefont {H{\"a}nggi},\ and\ \citenamefont
  {Talkner}}]{Grabert1980}%
  \BibitemOpen
  \bibfield  {author} {\bibinfo {author} {\bibfnamefont {H.}~\bibnamefont
  {Grabert}}, \bibinfo {author} {\bibfnamefont {P.}~\bibnamefont {H{\"a}nggi}},
  \ and\ \bibinfo {author} {\bibfnamefont {P.}~\bibnamefont {Talkner}},\ }\href
  {\doibase 10.1007/BF01011337} {\bibfield  {journal} {\bibinfo  {journal} {J.
  Stat. Phys.}\ }\textbf {\bibinfo {volume} {22}},\ \bibinfo {pages} {537}
  (\bibinfo {year} {1980})}\BibitemShut {NoStop}%
\bibitem [{\citenamefont {Grigolini}(1982)}]{Grigolini1982}%
  \BibitemOpen
  \bibfield  {author} {\bibinfo {author} {\bibfnamefont {P.}~\bibnamefont
  {Grigolini}},\ }\href {\doibase 10.1007/BF01008940} {\bibfield  {journal}
  {\bibinfo  {journal} {J. Stat. Phys.}\ }\textbf {\bibinfo {volume} {27}},\
  \bibinfo {pages} {283} (\bibinfo {year} {1982})}\BibitemShut {NoStop}%
\bibitem [{\citenamefont {Kupferman}(2004)}]{Kupferman2004}%
  \BibitemOpen
  \bibfield  {author} {\bibinfo {author} {\bibfnamefont {R.}~\bibnamefont
  {Kupferman}},\ }\href {\doibase 10.1023/B:JOSS.0000003113.22621.f0}
  {\bibfield  {journal} {\bibinfo  {journal} {J. Stat. Phys.}\ }\textbf
  {\bibinfo {volume} {114}},\ \bibinfo {pages} {291} (\bibinfo {year}
  {2004})}\BibitemShut {NoStop}%
\bibitem [{\citenamefont {Siegle}\ \emph {et~al.}(2010)\citenamefont {Siegle},
  \citenamefont {Goychuk}, \citenamefont {Talkner},\ and\ \citenamefont
  {H{\"a}nggi}}]{Siegle2010}%
  \BibitemOpen
  \bibfield  {author} {\bibinfo {author} {\bibfnamefont {P.}~\bibnamefont
  {Siegle}}, \bibinfo {author} {\bibfnamefont {I.}~\bibnamefont {Goychuk}},
  \bibinfo {author} {\bibfnamefont {P.}~\bibnamefont {Talkner}}, \ and\
  \bibinfo {author} {\bibfnamefont {P.}~\bibnamefont {H{\"a}nggi}},\ }\href
  {http://link.aps.org/doi/10.1103/PhysRevE.81.011136} {\bibfield  {journal}
  {\bibinfo  {journal} {Phys. Rev. E}\ }\textbf {\bibinfo {volume} {81}},\
  \bibinfo {pages} {011136} (\bibinfo {year} {2010})}\BibitemShut {NoStop}%
\bibitem [{\citenamefont {Cox}\ and\ \citenamefont {Miller}(1977)}]{Cox1977}%
  \BibitemOpen
  \bibfield  {author} {\bibinfo {author} {\bibfnamefont {D.}~\bibnamefont
  {Cox}}\ and\ \bibinfo {author} {\bibfnamefont {H.}~\bibnamefont {Miller}},\
  }\href@noop {} {\emph {\bibinfo {title} {{The Theory of Stochastic
  Processes}}}}\ (\bibinfo  {publisher} {Chapman and Hall/CRC},\ \bibinfo
  {year} {1977})\BibitemShut {NoStop}%
\bibitem [{\citenamefont {Breuer}(2004)}]{Breuer2004}%
  \BibitemOpen
  \bibfield  {author} {\bibinfo {author} {\bibfnamefont {H.-P.}\ \bibnamefont
  {Breuer}},\ }\href {https://link.aps.org/doi/10.1103/PhysRevA.70.012106}
  {\bibfield  {journal} {\bibinfo  {journal} {Phys. Rev. A}\ }\textbf {\bibinfo
  {volume} {70}},\ \bibinfo {pages} {012106} (\bibinfo {year}
  {2004})}\BibitemShut {NoStop}%
\bibitem [{\citenamefont {Budini}(2013)}]{Budini2013}%
  \BibitemOpen
  \bibfield  {author} {\bibinfo {author} {\bibfnamefont {A.~A.}\ \bibnamefont
  {Budini}},\ }\href {https://link.aps.org/doi/10.1103/PhysRevA.88.032115}
  {\bibfield  {journal} {\bibinfo  {journal} {Phys. Rev. A}\ }\textbf {\bibinfo
  {volume} {88}},\ \bibinfo {pages} {032115} (\bibinfo {year}
  {2013})}\BibitemShut {NoStop}%
\bibitem [{\citenamefont {Kretschmer}\ \emph {et~al.}(2016)\citenamefont
  {Kretschmer}, \citenamefont {Luoma},\ and\ \citenamefont
  {Strunz}}]{Kretschmer2016}%
  \BibitemOpen
  \bibfield  {author} {\bibinfo {author} {\bibfnamefont {S.}~\bibnamefont
  {Kretschmer}}, \bibinfo {author} {\bibfnamefont {K.}~\bibnamefont {Luoma}}, \
  and\ \bibinfo {author} {\bibfnamefont {W.~T.}\ \bibnamefont {Strunz}},\
  }\href {https://link.aps.org/doi/10.1103/PhysRevA.94.012106} {\bibfield
  {journal} {\bibinfo  {journal} {Phys. Rev. A}\ }\textbf {\bibinfo {volume}
  {94}},\ \bibinfo {pages} {012106} (\bibinfo {year} {2016})}\BibitemShut
  {NoStop}%
\bibitem{Breuer2008}  H.-P. Breuer and  B. Vacchini, Phys. Rev. Lett.  \textbf{101}, 140402 (2008).
  \bibitem [{\citenamefont {Cox}(1955)}]{Cox1955}%
  \BibitemOpen
  \bibfield  {author} {\bibinfo {author} {\bibfnamefont {D.~R.}\ \bibnamefont
  {Cox}},\ }\href {\doibase 10.1017/S0305004100030437} {\bibfield  {journal}
  {\bibinfo  {journal} {Math. Proc. Cambridge}\ }\textbf {\bibinfo {volume}
  {51}},\ \bibinfo {pages} {433} (\bibinfo {year} {1955})}\BibitemShut
  {NoStop}%
\bibitem [{\citenamefont {Beylkin}\ and\ \citenamefont
  {Monz{\'o}n}(2005)}]{Beylkin2005}%
  \BibitemOpen
  \bibfield  {author} {\bibinfo {author} {\bibfnamefont {G.}~\bibnamefont
  {Beylkin}}\ and\ \bibinfo {author} {\bibfnamefont {L.}~\bibnamefont
  {Monz{\'o}n}},\ }\href {\doibase 10.1016/j.acha.2005.01.003} {\bibfield
  {journal} {\bibinfo  {journal} {Appl. Comput. Harmon. Anal.}\ }\textbf {\bibinfo
  {volume} {19}},\ \bibinfo {pages} {17} (\bibinfo {year} {2005})}\BibitemShut
  {NoStop}%
\bibitem [{\citenamefont {Levin}\ \emph {et~al.}(2008)\citenamefont {Levin},
  \citenamefont {Peres},\ and\ \citenamefont {Wilmer}}]{Levin2008}%
  \BibitemOpen
  \bibfield  {author} {\bibinfo {author} {\bibfnamefont {D.}~\bibnamefont
  {Levin}}, \bibinfo {author} {\bibfnamefont {Y.}~\bibnamefont {Peres}}, \ and\
  \bibinfo {author} {\bibfnamefont {E.}~\bibnamefont {Wilmer}},\ }\href@noop {}
  {\emph {\bibinfo {title} {{Markov Chains and Mixing Times}}}},\ \bibinfo
  {edition} {1st}\ ed.\ (\bibinfo  {publisher} {American Mathematical
  Society},\ \bibinfo {year} {2008})\BibitemShut {NoStop}%
\bibitem [{\citenamefont {Golub}\ and\ \citenamefont {{Van
  Loan}}(1996)}]{Golub1996}%
  \BibitemOpen
  \bibfield  {author} {\bibinfo {author} {\bibfnamefont {G.~H.}\ \bibnamefont
  {Golub}}\ and\ \bibinfo {author} {\bibfnamefont {C.~F.}\ \bibnamefont {{Van
  Loan}}},\ }\href@noop {} {\emph {\bibinfo {title} {{Matrix Computations}}}},\
  \bibinfo {edition} {3rd}\ ed.,\ Vol.~\bibinfo {volume} {3}\ (\bibinfo
  {publisher} {The Johns Hopkins University Press, Baltimore},\ \bibinfo {year}
  {1996})\BibitemShut {NoStop}%
\bibitem [{\citenamefont {Vacchini}(2012)}]{Vacchini2012}%
  \BibitemOpen
  \bibfield  {author} {\bibinfo {author} {\bibfnamefont {B.}~\bibnamefont
  {Vacchini}},\ }\href {http://stacks.iop.org/0953-4075/45/i=15/a=154007}
  {\bibfield  {journal} {\bibinfo  {journal} {J. Phys. B}\ }\textbf {\bibinfo {volume} {45}},\ \bibinfo
  {pages} {154007} (\bibinfo {year} {2012})}\BibitemShut {NoStop}%
\bibitem [{\citenamefont {Nielsen}\ and\ \citenamefont
  {Chuang}(2010)}]{Nielsen2010}%
  \BibitemOpen
  \bibfield  {author} {\bibinfo {author} {\bibfnamefont {M.~A.}\ \bibnamefont
  {Nielsen}}\ and\ \bibinfo {author} {\bibfnamefont {I.~L.}\ \bibnamefont
  {Chuang}},\ }\href@noop {} {\emph {\bibinfo {title} {{Quantum Computation and
  Quantum Information}}}}\ (\bibinfo  {publisher} {Cambridge University Press,
  NY},\ \bibinfo {year} {2010})\BibitemShut {NoStop}%
\bibitem [{\citenamefont {Grifoni}\ \emph {et~al.}(1996)\citenamefont
  {Grifoni}, \citenamefont {Sassetti},\ and\ \citenamefont
  {Weiss}}]{Grifoni1996}%
  \BibitemOpen
\bibfield  {journal} {  }\bibfield  {author} {\bibinfo {author} {\bibfnamefont
  {M.}~\bibnamefont {Grifoni}}, \bibinfo {author} {\bibfnamefont
  {M.}~\bibnamefont {Sassetti}}, \ and\ \bibinfo {author} {\bibfnamefont
  {U.}~\bibnamefont {Weiss}},\ }\href {\doibase 10.1103/PhysRevE.53.R2033}
  {\bibfield  {journal} {\bibinfo  {journal} {Phys. Rev. E}\ }\textbf {\bibinfo
  {volume} {53}},\ \bibinfo {pages} {R2033} (\bibinfo {year}
  {1996})}\BibitemShut {NoStop}%
\bibitem [{\citenamefont {Pershin}\ and\ \citenamefont {{Di
  Ventra}}(2011)}]{Pershin2011}%
  \BibitemOpen
\bibfield  {journal} {  }\bibfield  {author} {\bibinfo {author} {\bibfnamefont
  {Y.~V.}\ \bibnamefont {Pershin}}\ and\ \bibinfo {author} {\bibfnamefont
  {M.}~\bibnamefont {{Di Ventra}}},\ } \href {\doibase
  10.1080/00018732.2010.544961} {\bibfield  {journal} {\bibinfo  {journal}
  {Adv. Phys.}\ }\textbf {\bibinfo {volume} {60}},\ \bibinfo {pages} {145}
  (\bibinfo {year} {2011})}\BibitemShut {NoStop}%
\bibitem [{\citenamefont {Driscoll}\ \emph {et~al.}(2009)\citenamefont
  {Driscoll}, \citenamefont {Kim}, \citenamefont {Chae}, \citenamefont {Kim},
  \citenamefont {Lee}, \citenamefont {Jokerst}, \citenamefont {Palit},
  \citenamefont {Smith}, \citenamefont {{Di Ventra}},\ and\ \citenamefont
  {Basov}}]{Driscoll2009}%
  \BibitemOpen
  \bibfield  {author} {\bibinfo {author} {\bibfnamefont {T.}~\bibnamefont
  {Driscoll}}, \bibinfo {author} {\bibfnamefont {H.-T.}\ \bibnamefont {Kim}},
  \bibinfo {author} {\bibfnamefont {B.-G.}\ \bibnamefont {Chae}}, \bibinfo
  {author} {\bibfnamefont {B.-J.}\ \bibnamefont {Kim}}, \bibinfo {author}
  {\bibfnamefont {Y.-W.}\ \bibnamefont {Lee}}, \bibinfo {author} {\bibfnamefont
  {N.~M.}\ \bibnamefont {Jokerst}}, \bibinfo {author} {\bibfnamefont
  {S.}~\bibnamefont {Palit}}, \bibinfo {author} {\bibfnamefont {D.~R.}\
  \bibnamefont {Smith}}, \bibinfo {author} {\bibfnamefont {M.}~\bibnamefont
  {{Di Ventra}}}, \ and\ \bibinfo {author} {\bibfnamefont {D.~N.}\ \bibnamefont
  {Basov}},\ }\href
  {http://science.sciencemag.org/content/325/5947/1518.abstract} {\bibfield
  {journal} {\bibinfo  {journal} {Science}\ }\textbf {\bibinfo {volume}
  {325}},\ \bibinfo {pages} {1518} (\bibinfo {year} {2009})}\BibitemShut
  {NoStop}%
\bibitem [{\citenamefont {Zheng}\ \emph {et~al.}(2013)\citenamefont {Zheng},
  \citenamefont {Yan},\ and\ \citenamefont {{Di Ventra}}}]{Zheng2013}%
  \BibitemOpen
  \bibfield  {author} {\bibinfo {author} {\bibfnamefont {X.}~\bibnamefont
  {Zheng}}, \bibinfo {author} {\bibfnamefont {Y.}~\bibnamefont {Yan}}, \ and\
  \bibinfo {author} {\bibfnamefont {M.}~\bibnamefont {{Di Ventra}}},\ }\href
  {https://link.aps.org/doi/10.1103/PhysRevLett.111.086601} {\bibfield
  {journal} {\bibinfo  {journal} {Phys. Rev. Lett.}\ }\textbf {\bibinfo
  {volume} {111}},\ \bibinfo {pages} {086601} (\bibinfo {year}
  {2013})}\BibitemShut {NoStop}%
\end{thebibliography}

%

\end{document}